\documentclass[]{aa}  

\usepackage{graphicx}
\usepackage{txfonts}
\usepackage{natbib}
\usepackage{deluxetable}
\begin{document}

   \title{Multi-wavelength study of the disk around the very 
low-mass star Par-Lup3-4\thanks{Based on observations obtained at the European Southern Observatory using the Very Large Telescope in Cerro Paranal, Chile, under programs 381.C-0283(A) and 71.C-0429(B).}}
   \author{N. Hu\'elamo\inst{1}
          \and
          H. Bouy\inst{1}
          \and
          C. Pinte\inst{2}
          \and
          F. M\'enard\inst{2}
          \and   
          G. Duch\^ene\inst{3,2}
          \and
          F. Comer\'on\inst{4}       
          \and
           M. Fern\'andez\inst{5}
          \and
          D. Barrado\inst{6,1} 
          \and
          A. Bayo\inst{7}
          \and 
          I. de Gregorio-Monsalvo\inst{7,8}
          \and
          J. Olofsson\inst{9}       
 }
   \offprints{N. Hu\'elamo}
   \institute{
  Centro de Astrobiolog\'{\i}a (INTA-CSIC); LAEFF, P.O. Box 78, E-28691 Villanueva de la 
   Ca\~nada, Spain\\
   \email{nhuelamo@cab.inta-csic.es}
      \and
   Laboratoire d'Astrophysique, Observatoire de Grenoble, 
   BP 53, F-38041 Grenoble, Cedex 9, France
   \and     
   Astronomy Dpt., 601 Campbell Hall, UC Berkeley, CA 94720 Berkeley, USA
   \and
     European Southern Observatory, Karl-Schwarzschild-Stra\ss e 2,
     D-85748 Garching, Germany
   \and
    Instituto de Astrof\'{\i}sica de Andaluc\'{\i}a, CSIC, Apdo. 3004,
    E-18080, Granada, Spain
    \and
    Calar Alto Observatory, Centro Astron\'omico Hispano Alem\'an, Almer\'{\i}a, Spain
    \and
    European Southern Observatory, Alonso de C\'ordova 3107, Casilla 19, 
    Vitacura, Santiago, Chile
    \and
    Atacama Large Millimeter/Submillimeter Array, Joint ALMA Office,
    Av. Apoquindo 3650, Piso 18, Las Condes, Santiago, Chile.    
    \and
    Max-Planck-Institut f\"ur Astronomie, Koenigstuhl 17, D-69117 Heidelberg, Germany    }
   \date{Received; accepted}
    
   \abstract {Par-Lup3-4 is a very low-mass star (spectral type M5) in
the Lupus~III star-forming region. It shows spectroscopic
evidence of accretion and mass-loss. In the
optical and near-infrared, the object is underluminous by
$\approx$4~mag when compared to objects of similar mass in the same
association. 
 }
{ The aim of this work is to characterize the circumstellar environment
of  Par-Lup3-4 to better understand the origin of its underluminosity. 
 }
{We have analyzed high angular resolution near-IR observations and searched for extended emission
from a disk and/or an envelope.
 We have studied the spectral energy distribution (SED) of the target 
 from the optical to the sub-millimeter regime, and 
 compared it to a grid of radiative transfer models of 
 circumstellar disks. Since the target is strongly variable, we modeled two 
 different near-infrared datasets.
 }
{
The SED of Par-Lup3-4 resembles that of objects with edge-on disks seen in scattered light, that is,
a double peaked-SED and a dip at $\sim$10 $\mu$m. 
The diffraction-limited infrared observations do not show obvious extended emission, allowing us to put an upper limit of $\sim$20\,AU
to the disk outer radius.  Par-Lup3-4 is probably in a Class II (rather than a Class I) evolutionary stage, which is indicated by the lack of extended 
emission together with the non detection of a strong 9.8$\mu$m silicate in absorption. This last feature is indeed seen in emission.
We fitted the whole SED of Par-Lup3-4 with a single disk model. Our modeling predicts a disk inclination of 81\degr$\pm$6\degr, 
which agrees well with previous estimates, and provides a natural explanation for the under-luminosity of the target.
The detection of the silicate feature in emission at such a high inclination
might be related to a more complex disk structure (e.g. asymmetries, inhomogeneities) than the one assumed here.
Our analysis allows us to put constraints on the disk inner radius,
R$_{\mathrm{in}}$ $\le$ 0.05 AU, which is very close to the dust
sublimation radius, and the maximum size of the dust grains,
a$_{\mathrm{max}}$ $\ge$ 10 $\mu$m, which indicates that
dust processing has already taken place in Par-Lup3-4. 
Some of the derived disk parameters vary depending on the modeled near-infrared 
data-set, which emphasizes the importance of taking variability into account when modeling the SED of
young stellar objects.
}
{}
\authorrunning{Hu\'elamo et al.}
\titlerunning{The circumstellar environment of the very low-mass star Par-Lup3-4}             
\maketitle
%

\section{Introduction}

A large number of very low-mass stars (VLMSs) and brown dwarfs (BDs)
in star-forming regions (SFR) are surrounded by circumstellar disks
\citep[see review by][]{Luhman2007}. Several studies have shown that these objects can undergo
a Classical T Tauri phase that is characterized by accretion and mass-loss
processes, like those observed in more massive pre-main sequence (PMS)
low-mass stars \citep[see  e.g][]{Luhman-1997,FC2001,Natta-2004,byn-2004b, mohanty-2005}.

Over the last few years, several authors have reported the detection
of young VLMS and BDs with a very peculiar phenomenology \citep[e.g.,
  LS-RCrA1, Par-lup-3-4, KPNO-Tau 12, 2MASSJ04381486+2611399, reported
  respectively by][]{FC2001, Comeron-2003, Luhman2003, Luhman2007}:
these objects show spectral features that ate characteristic of accretion, in
some cases of jets and outflows, and all of them are under-luminous
when compared to objects of similar spectral type and ages in the same
SFR.

\citet{FC2001} and \citet{Comeron-2003} have discussed three possible
scenarios to explain this under-luminosity: (i) accretion-modified
evolution, as described by e.g. \citet[][]{1997ApJ...475..770H,Baraffe2009}; (ii) embedded
Class~I sources extincted by a dusty envelope; (iii) Class~II objects whose disks are seen near
edge-on and block part of the stellar light. 


Par-Lup3-4 is a very low-mass star discovered in the course of a near-infrared survey in
the Lupus~III cloud \citep[source 0529.9-5737,][]{Nakajima-2000}.
It is an M5 star with a rich emission-line spectrum displaying features
commonly associated with accretion and outflow  \citep{Comeron-2003}. It is under-luminous
by almost 4\,mag when compared with other objects of similar spectral
type and age in the same star-forming region. The low luminosity
places the star in the 50\,Myr isochrone on the H-R diagram, that is,
much older than the age estimated for other members of the Lupus~III SFR
\citep[well below 5~Myr,][]{Comeron-2003}.

\citet{FC2005} presented deep optical imaging and spectroscopy of Par-Lup3-4
obtained with FORS1 and UVES at the Very Large Telescope.  
The H$_{\alpha}$ and narrow-band [\ion{S}{ii}]
images revealed extended emission associated with a jet, which can be
traced up to 4\farcs2 from the star. 
The faint jet, designated as
HH~600, is detected on each side of the central star, with the
southeastern component displaying a bright knot well detected in both
H$\alpha$ and [\ion{S}{ii}] as far as 1\farcs3 from the star.  These
images make of Par-Lup3-4 one of the least massive stellar objects known to excite a
jet.  The optical spectrum shows a double peak [\ion{S}{ii}] line
profile with the peaks separated by 40\,km/s, suggesting that the jet is seen at a small angle with respect
to the plane of the sky.  Assuming velocities for the emission knots
similar to those commonly found in jets of more massive young stellar
objects (T Tauri stars, 100-150~km/s), \citet{FC2005} have derived an
orientation of the jet between 8\degr and 12\degr\, (0\degr\, being the
plane of the sky), suggesting that the associated disk is seen nearly
edge-on.

The availability of new infrared and sub-millimeter data of Par-Lup3-4 gives us the opportunity of characterizing  
its circumstellar environment and study if its  under-luminosity can be explained by the 
presence of an edge-on disk or an envelope. In addition, we can
check the consistency  with the results derived using optical spectroscopy by \citet{FC2005}.
In this work, we present diffraction-limited  near-infrared observations of Par-Lup3-4, together with mid-IR spectroscopic and sub-millimeter data. We have used all these observations and archival or published data to build and model the spectral energy distribution (SED) of the target. In Sect.~2, we give an
overview of all the observational data. The main properties of the SED and its 
modeling are described in Sect.~3 and 4. The analysis  and main results
are described in  Sect.~5, while the main conclusions are
summarized in Sect.~6.


\section{Par-Lup3-4 observational data}

\subsection{High angular resolution near-IR observations of Par-Lup3-4}

\begin{table}
\caption{NACO/VLT observations}
\advance\tabcolsep by -2pt
\begin{tabular}{lllll}
\hline
Date &              Filter    &     Exp. Time     &     Seeing   &            Airmass            \\
         &                          &     [s]            &             ["]      &                                            \\ \hline 
22/06/2008  &  $K_s$           & 600            &          0.60-0.62     &     1.12  \\
22/06/2008  &  $NB\,2.12$   & 1800          &         0.65-0.62       &   1.10  \\
22/06/2008   & $L' $          & 1440           &          0.85-0.91      &    1.23  \\ \hline
\hline
\end{tabular}\label{naco_log}
\end{table}
\begin{figure}[h]
   \centering
   \resizebox{\hsize}{!}{\includegraphics[angle=0]{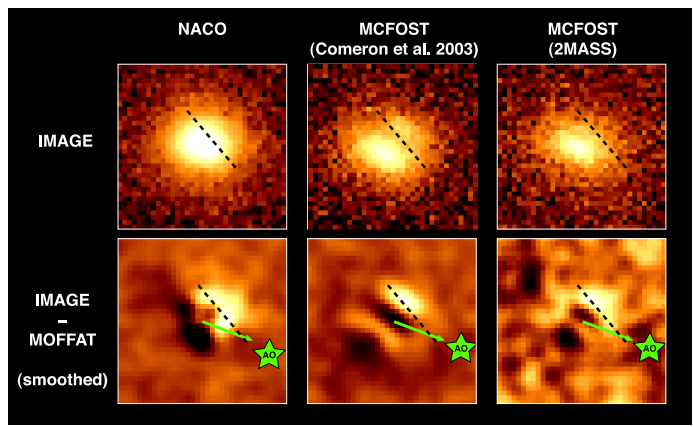}}
\caption{{\bf Left}: NACO/VLT $K_s$ image of Par-Lup3-4  (top) and the result of subtracting a Moffat function
 (bottom). The dashed line represents the expected orientation of the circumstellar disk 
 according to \citet{FC2005}, and the solid line the direction of the NAOS reference star. 
{\bf Middle and right panels}: MCFOST synthetic images of our two
  best models. We have fitted two datasets, C03 and 2MASS, which were
  acquired at different times and correspond to the most extreme
  (brightest and faintest) near-IR photometry of the target (see Sect.~4).  The
bottom panels show the residuals after PSF-subtraction. North is up
and East to the left.}
   \label{nacoima}
\end{figure}
Par-Lup3-4 was observed in service mode on 2008 June 22 with NAOS-CONICA (NACO), the Adaptive Optics (AO) facility  and near-IR camera at the Very Large Telescope (VLT). 
The object is too faint to be used as a reference star for the adaptive optics wavefront sensor, 
and a nearby star was used instead (Par-Lup3-3, $K_s$=9.54\,mag, located at 24\farcs4
southwest of the target).

The target was observed in the $K_s$ and $L'$ broad band filters to search
for extended emission from the disk (and envelope), and in the 2.12\,$\mu$m narrow
band filter, to search for H$_2$ (1-0 S\,(1)) emission from shocks
related to its jet.  The S27 camera was used for the {\em K$_s$} and {\em NB2.12}
observations and the L27 camera for the {\em L'} observations, leading to a
nominal pixel scale of $\sim$27\,mas/pix and a total field of view of
$\sim$27\arcsec $\times$ 27\arcsec.  Observations in the $K_s$ and $NB2.12$
filter were made with the infrared wavefront sensor and the N90C10
dichroic that sends 10\% of the light to the science camera and 90\%
to the wavefront sensor. The {\em L'} observations were carried out with the
near-IR wavefront sensor and the JHK dichroic that sends the JHK light
of the reference star to the wavefront sensor.

The {\em Ks} and {\em NB2.12} observations were made in jitter mode with 10 random
positions and a jitter width of 7".  The {\em L'} band observations were
made in jitter mode as well, with 48 random positions within a 7"
jitter box. The total exposure times add up to 600~s, 1800~s, and
1440~s in the {\em K$_s$}, {\em NB2.12}, and {\em L'} filters, respectively.  

The NACO data were processed with the {\em Eclipse} software
\citep{Devi1997} following the standard reduction steps, that is,
dark-subtraction, flat-fielding, bad pixel correction, sky subtraction, 
and image stacking. Table~\ref{naco_log} gives an overview of the
observational settings and properties.

Par-Lup3-4 is detected in the three NACO filters. The target is seen
slightly elongated in the direction of the reference star. This
elongation could be caused by the adaptive optics correction, given that the
wavefront sensing reference star was not the target itself and the PSF
is expected to suffer from anisoplanatism.  To study this elongation
and the presence of extended emission, we have performed
PSF-subtraction and analyzed the residuals.

Because no PSF star could be obtained, we fitted Par-Lup3-4 $K_s$-band
image with a MOFFAT function and subtracted it. The result is shown in
the left panels of Fig.~\ref{nacoima}. The PSF-subtracted image shows
asymmetric residuals, with a faint extended nebulosity in the northwest 
quadrant. The nebulosity is not aligned in the direction of the
reference star, but is closer to the expected position of the disk, at
a position angle of $\sim$ 40$\deg$ North to East, that is,
perpendicular to the jet \citep{FC2005}.  In any case, the signal-to-noise ratio 
of the dataset is not sufficient to conclude that the disk is spatially
resolved.  The {\em Ks}-band residuals allow us to derive an upper limit
for the disk angular radius of $\sim$0\farcs1.  For the
{\em L'} image, a MOFFAT fit on the NACO image shows that indeed no
significant residuals remain in the northwestern quadrant. This is
mostly owing to the much lower signal-to-noise ratio of the {\em L'} image
(115 instead of 600).

\subsection{Mid-infrared spectroscopy}

\begin{figure}[t]
   \centering
   \resizebox{\hsize}{!}{\includegraphics[angle=0]{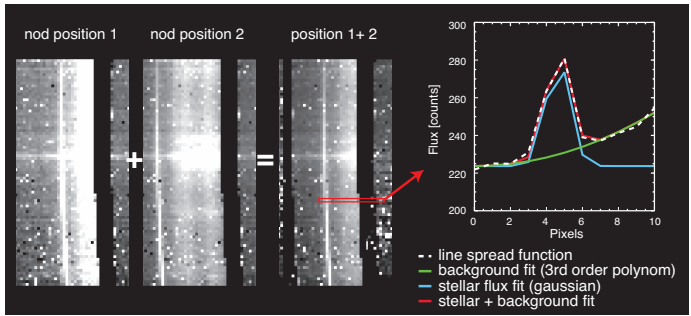}}
   \caption{{\it Spitzer} IRS 2D spectrum of Par-Lup-3-4 on the long
     wavelength range. The two nodding positions display a strong and variable background on
     the right hand side of the image that is not properly cancelled
     after subtraction. We therefore combined the two images, and extracted the spectrum
     by fitting a Gaussian + 3rd order polynomial function to the
     spread line function at each line, as shown in the right panel
     for a given line.}
   \label{spectrum}
\end{figure}

Par-Lup3-4 was observed with the {\it Spitzer Infrared Spectrograph}
\citep[IRS, ][]{2004SPIE.5487...62H} as part of program 40134
(P.I. Luhman) during cycle 7. The source was observed in staring mode
at low resolution at both short and long wavelengths.  We retrieved
the public data and first processed them using the \emph{c2d} package,
which is fully described in \citet[][]{Spizter_IRS_Reduction}. Two
different extraction methods are available: full aperture extraction
or optimal PSF extraction. The second method is less sensitive to bad
pixels or bad data samples than the first method which provides spectra
with fewer spikes. We used the spectrum and corresponding uncertainties
obtained with the PSF extraction method. The \emph{c2d} pipeline
furthermore corrects for possible pointing errors that can lead to
important offsets between different modules.

The short wavelength spectrum ranging from 5 to $\simeq$~14$\mu$m agrees
well with the overlapping IRAC photometry, but the long
wavelength spectrum was clearly problematic with a 14~$\mu$m flux
that was inconsistent with the overlapping short wavelength spectrum, and a
24~$\mu$m flux inconsistent with the MIPS photometry.

A careful inspection of the 2-D spectra reveals that the short
wavelength spectrum looks relatively clean and normal, while the long
wavelength 2-D spectrum displays a strong and spatially variable
background that does not cancel after subtracting the two nodding
positions, which explains the problematic extraction with the automated
\emph{c2d} pipeline (see Fig.~\ref{spectrum}). A different extraction
strategy was required, and we proceeded as follows: the two nodding
positions were aligned and combined rather than subtracted one from the
other. Obvious rogue pixels close to the stellar spectrum were cleaned
manually with the IRSCLEAN package within the Interactive Data
Language (IDL). Then, a synthetic line spread function made of a
Gaussian (for the source's signal) plus a third order polynomial
function (for the background) was fitted to each line of the 2D
spectrum. The corresponding synthetic background was then subtracted
from the original image, which was subsequently processed following the
recommended procedure with the {\it Spitzer} IRS Custom Extraction (SPICE)
software. The whole procedure is presented schematically in
Fig.~\ref{spectrum}. Unfortunately, uncertainties cannot be easily propagated
during this complex process. While it is difficult to assess
the reliability of the resulting spectrum, we note that it very well matches
the short wavelength spectrum on one side and the MIPS
24~$\mu$m photometry on the other side. Since we are not interested in
the detailed study of the different features and lines present in the
spectrum but in the general shape and slope of the spectrum, we
finally smoothed the spectrum using a Gaussian kernel with a width of 5
pixels.


\begin{figure}
   \centering
   \resizebox{\hsize}{!}{\includegraphics[angle=0]{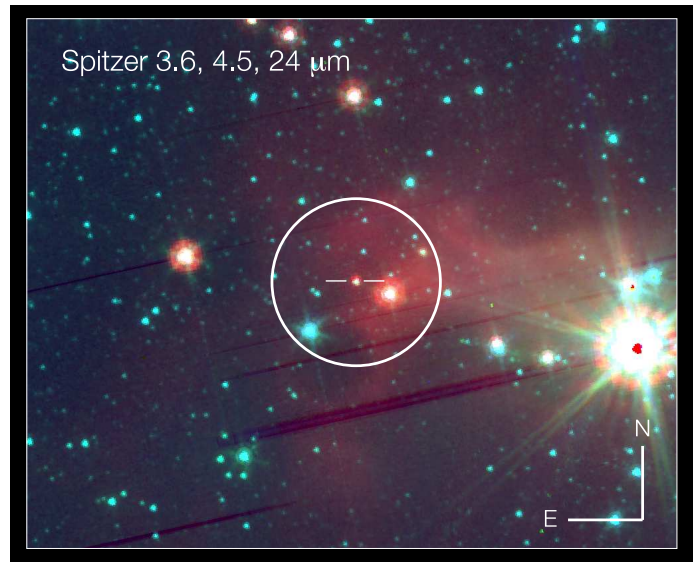}}
   \caption{Three-color {\em Spitzer} image of Par-Lup3-4 at 3.6 (blue), 4.5
     (green) and 24~$\mu$m (red).
     The circle centered on Par-Lup3-4
     has 1\arcmin\, radius and contains another Lupus~III member, Par-Lup3-3,
     $\sim$0\farcm5~SW from the target. Note the extremely red
     color or Par-Lup3-4 compared with other sources in the
     field of view. }
     \label{Spitzer_ima} 
     \end{figure}
   
\subsection{Submilimeter observations with APEX/LABOCA}

The Lupus~III region was observed at 870$\mu$m with LABOCA, the
bolometer array installed on the Atacama Pathfinder EXperiment
(APEX\footnote{This work is partially based on observations with the
  APEX telescope. APEX is a collaboration between the
  Max-Plank-Institute fur Radioastronomie, the European Southern
  Observatory, and the Onsala Space Observatory.}) telescope. The
angular resolution of each bolometer is 19.5$\pm$1$''$, and the total
field of view (FOV) of the array is 11.4$'$$\times$11.4$'$

We retrieved the data from the ESO archive. Observations were
performed on 2007 August 01 and 02 during the ESO program 079.F-9302A
(P.I. Hatchell), and on 2008 June 01, 02, 03, and 04 during the
program 081.F-9318A (P.I. Hatchell). The total on-source integration
time was $\simeq$29 hours. A spiral raster mapping was used as
observing pattern, providing a fully sampled and homogeneously covered
map in an area of diameter $\simeq$25$'$$\times$20$'$, centered at
$\alpha$ = 16$^{h}$09$^{m}$00.00$^{s}$, $\delta$ = -39$^{\rm
  o}$06$'$30.0$''$(J2000.0).

The data were acquired under relatively good weather conditions (zenith opacity 
values at 870\,$\mu$m ranged from 0.25 to 0.44, with a
mean value of 0.31).  The telescope pointing was checked every hour,
finding an rms pointing accuracy of $\simeq$2$''$. Focus settings were
checked once per night and during the sunset. Calibration was
performed using observations of Mars, Saturn, and Jupiter, as well as
secondary calibrators. The absolute flux calibration uncertainty is
estimated to be $\simeq$ 11$\%$.

We reduced the data with the BOlometer Array Analysis Software (BoA)
and the MiniCRUSH program (see \citealt{Kov08}). The pre-processing
steps consisted of flagging dead or cross-talking channels, frames
with too high telescope accelerations and with unsuitable mapping
speed, as well as temperature drift correction using two blind
bolometers. The data reduction process included flat-fielding, opacity
correction, calibration, correlated noise removal, and de-spiking.
Every scan was visually inspected to identify and discard corrupted
data. Finally, the individual maps were coadded and smoothed to a
final angular resolution of ~27\farcs6. We optimized the data reduction to
recover faint and point-like sources. As a result, we obtained a
1$\sigma$ point source sensitivity of 4.5 mJy.



Par-Lup3-4 is not detected in the LABOCA observations,  and we derive a
3-$\sigma$ upper limit for its 870$\mu$m continuum emission of 13.5 \,mJy.

\subsection{Optical, near-infrared, and mid-IR photometry}

\begin{center}
\begin{table}
\caption{Par-Lup3-4  Photometry}
\label{photometry}
\advance\tabcolsep by -2pt
\begin{tabular}{llll}\hline\hline
 Filter               & Magnitude      & Epoch       & Reference     \\
                      & [mag]          &             & \\ 
\hline\noalign{\smallskip}
$V $                    & 21.9$\pm$0.3               & 2002-03-22  & \citet{Comeron-2003} \\
$R_{\rm c}$   & 19.78$\pm$0.15          & 2002-03-22  & \citet{Comeron-2003} \\
$R_{\rm c}$   & 18.97$\pm$0.59 & 2003-04-25  & \citet{merin2008} \\
$I_{\rm c}$    & 18.3$\pm$0.1           & 2002-03-22  & \citet{Comeron-2003} \\
$I_{\rm c}$    & 18.18$\pm$0.05          & 2003-04-25  & \citet{merin2008} \\
$J $                     & 15.70$\pm$0.03  & 1998-03-12  & \citet{Nakajima-2000} \\
$J $                     & 16.04$\pm$0.21	& 1999-06-04  & DENIS \\
$J $                     & 15.46$\pm$0.06  & 2000-03-30  & 2MASS  \\
$J$                      & 16.42$\pm$0.05  & 2002-07-20  & \citet{Comeron-2003} \\
$H $                    & 14.72$\pm$0.02  & 1998-03-12  & \citet{Nakajima-2000} \\  
$H $                    & 14.26$\pm$0.04  & 2000-03-30  & 2MASS  \\   
$H$                     & 15.09$\pm$0.05  & 2002-07-20  & \citet{Comeron-2003} \\  
$K_s$                    & 13.70$\pm$0.01  & 1998-03-12  & \citet{Nakajima-2000} \\  
$K_s$                    & 13.58$\pm$0.19  & 1999-06-04  & DENIS \\
$K_s$                    & 13.30$\pm$0.04  & 2000-03-30  & 2MASS  \\                 
$K_s$                    & 13.82$\pm$0.10  & 2002-07-20  & \citet{Comeron-2003} \\ 
\hline\noalign{\smallskip}
Filter                & Flux            & Epoch       & Reference \\
                      & [mJy]          &             & \\ 
\hline  \noalign{\smallskip}
IRAC1 (3.6~$\mu$m)    & 3.11$\pm$0.05   & 2004-08-24  & \citet{chapman2007}\\ 
IRAC2 (4.5~$\mu$m)    & 3.25$\pm$0.04   & 2004-08-24  & \citet{chapman2007} \\ 
IRAC3 (5.8~$\mu$m)    & 2.34$\pm$0.04   & 2004-08-24  & \citet{chapman2007} \\ 
IRAC4 (8.0~$\mu$m)    & 1.74$\pm$0.04   & 2004-08-24  & \citet{chapman2007} \\ 
MIPS1 (24~$\mu$m)     & 27.9$\pm$0.3     & 2004-08-24  & \citet{chapman2007} \\  
MIPS2 (70~$\mu$m)     & 492.0$\pm$53.6  &             & \citet{merin2008} \\
\hline
\end{tabular}
\end{table}
\end{center}

Par-Lup3-4 has been observed in different filters and epochs.
Table \ref{photometry} gives an overview of the compiled photometry of the
source in different filters.

 \citet{Comeron-2003} presented simultaneous optical $V$,
$R$, and $I_{\rm c}$ photometry. \citet{merin2008} have provided
$R_{\rm c}$$I_{\rm c}$ photometry. The target shows variability in  the $R_{\rm c}$ 
band despite the large photometric errors.
In this paper, we will only consider the simultaneous optical observations
by \citet{Comeron-2003}.

Four independent studies report near-infrared (near-IR) measurements: the DENIS and 2MASS surveys
\citep[][]{DENIS, 2MASS}, \citet{Nakajima-2000} and
\citet{Comeron-2003}.  Par-Lup3-4 is variable at near-IR wavelengths as
already reported by \citet{Comeron-2003}. Such a strong variability ($\sim$0.5-1 mag) is
not uncommon among T Tauri stars, and
it can be attributed to a combination of changes in the accretion process and disk
properties, as reported in previous studies of these objects
\citep[e.g.][]{Mendoza1968, Appen1989,Skru1996,Eiroa2002,2007AJ....133..845W}.

Par-Lup3-4 was observed with the {\it Spitzer Space Telescope} on 2004
August 24 in the course of the  {\em Cores to Disks} (\emph{c2d}) Legacy Program
\citep[Program 175, P.I. Evans, ][ and
  Fig.~\ref{Spitzer_ima}]{chapman2007}. It is reported in the
\emph{c2d} catalog in the four IRAC filters (3.6, 4.5, 5.8 and
8.0~$\mu$m) and with MIPS at 24 and 70~$\mu$m
\citep[][]{chapman2007,merin2008}.  A mid-IR image of the region is displayed in Fig.~\ref{Spitzer_ima}.


\section{The SED of Par-Lup3-4}

\begin{figure}[t]
   \centering
   \resizebox{\hsize}{!}{\includegraphics{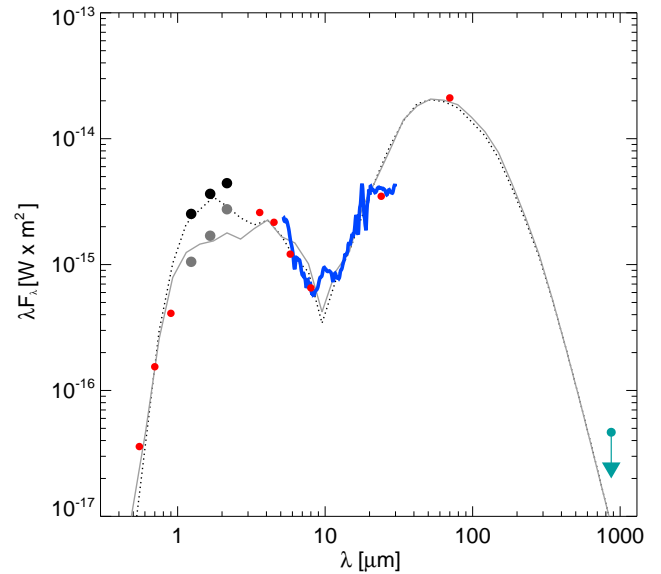}}
   \caption{Observed SED of Par-Lup3-4. The circles show all photometric measurements.
  For the near-IR regime, the black circles represent 2MASS data, and the grey
  ones C03 data. The strong  variability is responsible  for the dispersion of the measurements 
  taken at these wavelengths. The {\it Spitzer} IRS spectrum is shown in blue color. 
  The best-fit models from our Bayesian analysis are represented
  by dotted (2MASS) and solid (C03)  lines.
    \label{sed_parlup34} }
   \end{figure}

The SED of Par-Lup3-4 is displayed in Fig.~\ref{sed_parlup34}.  We
have overplotted the two most extreme near-IR datasets given in
Table~\ref{photometry} to show its strong variability.  The most
remarkable feature of the SED is its overall shape: two-peaks at near-
and mid-IR wavelengths and a dip around 8$\mu$m. This shape is typical
of edge-on disks obscuring the central star
\citep[e.g.][]{Stapelfeldt99, 2002ApJ...567.1183W,duchene2010}.  Because the
disk is optically thick at short wavelengths, the visible and
near-IR stellar light is blocked and the emission at these wavelengths
is dominated by scattered light. At longer wavelengths ($>$24~$\mu$m)
the dust emission penetrates the mostly optically thin disk, resulting
in a steep positive slope.

Since the object is observed in scattered light at shorter
wavelengths, we cannot derive a visual extinction using its
colors. This and the double-peak SED prevent us from
using the classical SED indexes that combine near and mid-IR data
\citep[e.g.][]{Lada1984, Luhman2008} to shed light on its evolutionary
stage.

The {\it Spitzer} IRS spectrum allows us a more detailed study of the
SED, at least from 5 to 15 microns, where the signal is properly
background-corrected.  The most interesting feature in this range is
the silicate band at $\sim$9.8 micron, which can provide information
about the evolutionary status of the object, the disk inclination, and
the properties of the dust.

\begin{figure}
   \centering
   \resizebox{\hsize}{!}{\includegraphics{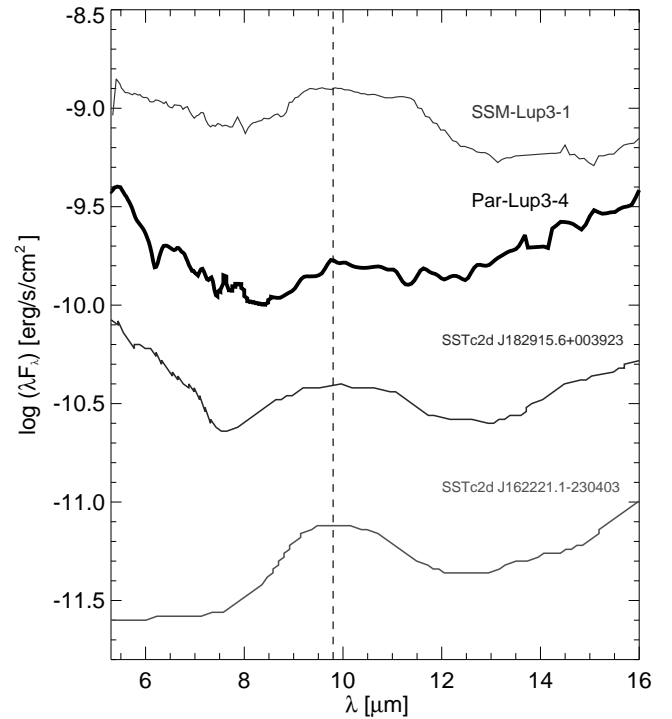}}
   \caption{{\it Spitzer} IRS spectrum of Par-Lup3-4 showing the silicate feature at 9.8$\mu$m (dashed line). 
   The feature is not detected in absorption  and points toward  an evolutionary stage closer to Class~II than Class~I.
   For comparison, we have overplotted the IRS spectrum of  three stellar 
   objects with disks at high inclinations and the silicate feature in emission:  
   SSM-Lup3-1 \citep{merin2007}, SSTc2d J182915.6+003923 and SSTc2d J162221.1-230403 \citep{merin2010}.
      \label{silicates}}
   
   \end{figure}

Figure~\ref{silicates} shows the SL IRS spectrum of Par-Lup3-4.  First,
note that we do not detect a strong and deep silicate
feature in absorption, as expected for Class~I objects surrounded by
thick envelopes \citep[e.g.][]{Watson2004,Furlan2008}.  A strong
CO$_{2}$ absorption band at 15.2 $\mu$m is not detected either, although
the problems with the background subtraction at those wavelengths
prevent us from drawing firm conclusions about this feature.  The
mid-IR spectrum points toward an evolutionary stage closer to
Class~II than Class~I, which is consistent with the lack of extended
emission from an envelope in the NACO/$K_s$ observations.

 The shape and strength of the silicate feature strongly depends on
 the disk inclination and the presence of small grains in the surface
 of the disk.  For edge-on disks, the study of the silicate
 feature is more complicated because the SED normally
 shows a dip at 10 microns, which is not related to the feature itself,
 but to the high inclination of the disk \citep[see e.g.][]{Ponto2007}. For Par-Lup3-4, the
 spectrum slope is inverted between 7$\mu$m and 14 $\mu$m and displays
   a dip at 8.5$\mu$m.     A flux enhancement is observed between
   9.8$\mu$m and 11$\mu$m (Fig. ~\ref{silicates}) and,
 although not expected in  disks at high inclinations, seems to be related to the silicate feature in
 emission. Indeed, \citet{merin2010} have recently presented several 
 close to edge-on disks that also show this feature in emission. Two of them are included in Fig. \ref{silicates} for comparison. A third stellar object from the same star-forming region and with an estimated disk inclination of $i\geq60$\,\degr, SSM-Lup3-1, has also been overplotted for comparison \citep{merin2007}.
The shape of the Par-Lup3-4 spectrum resembles that of SSTc2d J182915.6+003923 (see Fig.~\ref{silicates}), a K7-type star in
Serpens with a disk inclination of $i = 81.4\,\degr$ \citep{merin2010}.  Understanding the dust properties of Par-Lup3-4 (e.g. grain size distribution, crystalline vs amorphous species)   would require a higher quality spectrum and a more  detailed analysis, including a more refined grid of models for the continuum than those presented in this work, and also a careful modeling
 of the mineralogy. Such an analysis goes beyond the scope of this
 paper, which aims at understanding the underluminosity of the target.

\section{SED modeling}

We have modeled the SED of Par-Lup3-4 in order to characterize its
circumstellar environment.  As explained in Sect.~4, the IRS
spectrum allows us to rule out the presence of a thick envelope. We
therefore neglect the contribution of a possible envelope remnant, and
model the SED assuming that the target is surrounded only by a disk.

Because several epochs are available for the near-IR photometry, and as the
source displays a significant variability in this wavelength range, we
have modeled two datasets: the faintest and brightest sets of near-IR
photometry, from \citet{Comeron-2003} and 2MASS, respectively.  In
both cases, the solution is certainly not unique, but constraints can
be drawn on some parameters.  The lack of additional and simultaneous
epoch observations in the other wavelength ranges prevents us
from studying the impact of optical (which is commonly observed in
young late-type stars) or mid-IR variability \citep[which may be also
  common in these objects, e.g.][]{Robb2005, MC2009} in our best model
solutions.

\subsection{MCFOST radiative transfer models}

SED calculations were performed using the 3D radiative transfer code
MCFOST \citep{Pinte2006}, which includes multiple scattering, passive
dust heating, and continuum thermal
re-emission.The model consists of a central star surrounded by a
parametric disk described by power-laws.

\noindent{\emph{-- Disk structure: }} The models consider a flared
density structure with a Gaussian vertical profile, $\rho(r,z) =
\rho_0(r)\,\exp(-z^2/2\,h(r)^2)$, and power-law distributions for the
surface density, $\Sigma(r) = \Sigma_0\,(r/r_0)^{\alpha}$, and the scale
height, $ h(r) = h_0\, (r/r_0)^{\beta}$, where $r$ is the radial
coordinate in the equatorial plane, and 
$h_0$ the scale height at the reference
radius $r_0 =100$ AU. These assumptions lead to a general expression
for the density at any point in the disk:
\begin{equation}
  \rho(r,z) = \rho_0\,\left(\frac{r}{r_0}\right)^{\alpha-\beta}\,
\exp\left(-\frac{1}{2}\left(\frac{z}{h(r)}\right)^2\right).
\end{equation}

The disk extends from an inner cylindrical radius $r_{in}$ to an outer
limit $r_{out}$. The radiative transfer calculations are performed in
full 3D, although the disk structure is axisymmetric (2D).
\vspace{0.3cm}

\noindent{\emph{-- Grain properties: }} The models consider homogeneous
porous spherical grains and use the dielectric constants described by
\cite{Mathis89} in their model A, which follows interstellar-like relative abundance of elements.  
The grain sizes are distributed according to the
power-law $dn(a) \propto a^{-3.7}\,da$, with $a_{\mathrm{min}}$ and
$a_{\mathrm{max}}$ the minimum and maximum sizes of grains, discussed below.  
The interstellar values from \cite{Mathis89} are $a_{\mathrm{min}}=0.005$~$\mu$m
 and $a_{\mathrm{max}}=0.9~\mu$m. The mean grain density is $0.5$
g.cm$^{-3}$. In this work, $a_{\mathrm{max}}$ is considered as a free
parameter in an attempt to constrain the amount of dust processing 
in the disk. Extinction and scattering opacities, scattering
phase functions and Mueller matrices are calculated using the Mie theory. 
Finally, note that we have not tried to vary the grain composition to fit the silicate feature observed in the IRS
spectrum.

\subsection{Stellar parameters}

The distance to the Lupus SFR has been derived in different works
and show values that can  range between 100-300\,pc  \citep[see extensive discussion by ][]{ComeronF2008}.  
For this paper, we have adopted an average distance of 140\,pc to Par-Lup3-4.

For our analysis, we have used the NextGen \citep{1998A&A...337..403B}
synthetic photospheric spectrum of a star with T$_{\rm eff}$=3\,100~K
(as derived from optical spectroscopy by C03) and 1 Myr, which is the
estimated age of the Lupus~III association according to different
works \citep[e.g.][]{Comeron-2003,makarov2007,merin2008}.

The shape of the SED (and the modeling presented in the
analysis below) indicates that Par-Lup3-4 is seen in scattered light in the
optical and near-IR range.  As a consequence, the extinction toward
the central star and its luminosity (radius) cannot be reliably
derived with the optical spectrum or the optical and near-IR
colors. This value, however, will be determined by our modeling.

For the remainder of our analysis, we have considered Par-Lup3-4 to have a
stellar radius of 1.3~R$_{\odot}$ and a mass of 0.13$M_{\odot}$, as
predicted by the NextGen models for a 3\,100~K star at an age of
1\,Myr. To account for uncertainties in the distance and the main
properties of the central source (e.g. radius, age), a scaling factor
{\em q} (see Table 3) has been applied to the synthetic SED to match
the observed one.

\subsection{Disk parameters}

Table \ref{param_range} gives a summary of the parameter space covered
for the disk model. Because the target does not show an obvious extension in our NACO observations, 
we have fixed the outer disk radius to 20\,AU, which is a conservative value based on the 
residuals of the NACO images at a distance of 140\,pc. 

The grid adds up to a total of 409\,500 models on the IAC Condor\textregistered\, High Throughput
Computing facility and the UC Berkeley Clustered computing facility.
Although we have computed the SED only at the wavelengths at which we
have available photometry, the thermal equilibrium computation allows
for photons of all wavelengths to travel through the disk.

\begin{center}
\begin{table}
\advance\tabcolsep by -3pt
{\tiny
\caption{Parameter space covered by our analysis \label{param_range}}
\begin{tabular}{lllcll}\hline\hline
Parameter & Min & Max & Number & Sampling & Description\\
        &     &     & of values & &      \\
\hline
\multicolumn{5}{c}{MCFOST input parameters} \\
\hline
$a_{\rm min}$ [$\mu$m]  & 0.005  & 0.005 & 1 & fixed & minimum grain size \\
 $a_{\rm max}$ [$\mu$m]  & 1      & 100   & 5 & log   & maximum grain size \\
 $\alpha$               &  -1.5 & -0.5 & 3 & linear & surface density index \\
 M$_{\rm disk}$ [M$_{\sun}$] & 10$^{-7}$ & 10$^{-5}$ & 6 & log & disk dust mass \\
 h$_{0}$@100AU [AU]       & 10    & 30           & 6 & linear & scale height \\
 R$_{\rm in}$ [AU]      & 0.05   & 1.0          & 5 &log & inner radius \\
R$_{\rm out}$ [AU]    & 20    & 20           & 1 & fixed & outer radius \\
 $\beta$             & 1.05   & 1.35   & 5 & linear & flaring angle index \\
$i$ [\degr]           & 0      & 90  & 10 & linear in cos{\em i} & inclination \\
\hline
\hline
$A_{\rm V}$ [mag]       & 0           & 10  &           10 & & extinction \\
$q$                      & 0           & 2     &          5 &  & scaling factor \\
\hline
\end{tabular}
}
\end{table}

\end{center}

\begin{table}
\advance\tabcolsep by -2pt
{\tiny
\centering
\caption{Best-fit parameters  from the Bayesian analysis for the two near-IR datasets\label{bestmodel}}
\begin{tabular}{lcc}\hline\hline
               & 2MASS    & Comeron-2003\\
\hline
$a_{\rm max}$ [$\mu$m] $^1$                            & 100           & 100         \\  
$\alpha$ $^1$   & -0.5 & -0.5 \\
M$_{\rm disk}$ [10$^{-6}$~M$_{\odot}$] & 0.25            & 0.63  \\  
h$_{0}$@10~AU [AU]            & 1.28                   & 2.29  \\  
R$_{\rm in}$ [AU]                    & 0.05                   & 0.05 \\
$\beta$                                    & 1.275                   & 1.125 \\ 
$i$ [\degr]                   & 81                      &  81 \\ 
$A_\mathrm{V}$ [mag]          & 3.5                         & 2.0  \\ 
 $q$                & 0.7                     & 0.7 \\ \hline
reduced-$\chi^2$              & 19.9     & 13.4   \\
\hline
\end{tabular}

$^1$ Note that these two parameters are at the limit of the explored range of values.

}
\end{table}

\section{Analysis and main results}\label{analysis}

\subsection{Bayesian analysis}
 
We have performed a comparison of the synthetic SEDs with the
observations.  The best fit has been defined as that with the lowest
$\chi^{2}$ and was found by a complete exploration of the parameter
space within the ranges described in Table~\ref{param_range}.  Note
that based on our experience with the code, a conservative Monte
Carlo noise of 10\% has been included in the models.  Because there
are degeneracies between parameters, we have not attempted to refine
the best model, which is defined as the closest grid point (that is, a
local minimum in the parameter space).  Instead, we have performed a
Bayesian analysis using the reduced $\chi^{2}$ to get an estimate of
the validity range for each of the explored parameters
\citep[][]{Press1992, Lay1997, Pinte2007}.  The results from this
analysis provide the most probable disk parameters.

A relative probability exp(--$\chi^{2}$ /2) is calculated for each
individual model. The relative likelihood for each of the parameters
is obtained by adding the individual probabilities of all the models
with a given parameter value. In other words, these probabilities are
the result of a marginalization of the parameter space successively
over all dimensions and not a cut through the parameter space. In that
sense, they account for potential correlations and interplay between
parameters, and quantitative error bars can be extracted from them.

\begin{figure}
   \centering
   \resizebox{\hsize}{!}{\includegraphics{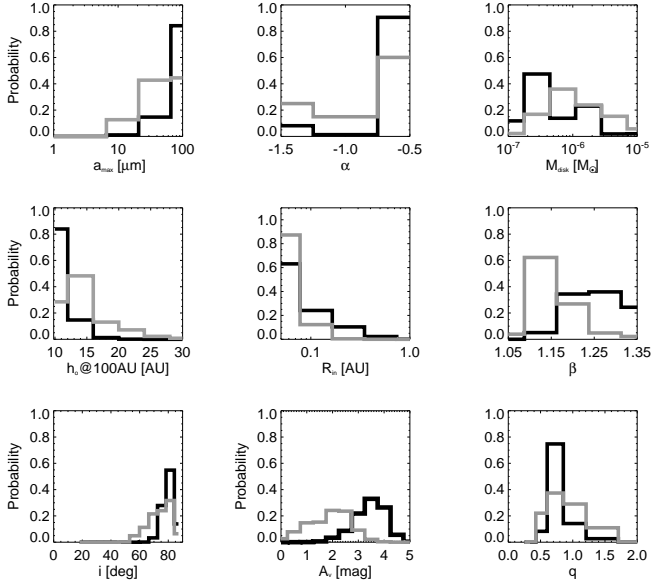} }
   \caption{Bayesian distribution of probability of the different
     disk parameters fitted by our model (see Table~\ref{bestmodel}) and described in 
     Table~\ref{param_range}. The black histograms represent the results obtained
     using the 2MASS near-IR photometry (brightest case) and the grey
     histograms the results obtained using the near-IR photometry of
     \citet{Comeron-2003} (faintest case). 
     \label{proba} }
   \end{figure}

The results from the Bayesian analysis are presented in
Fig.~\ref{proba}, where we display histograms with the probability for
different disk parameters.  We now discuss what constraints
on the properties of the disk can be extracted from this analysis.

\subsection{Disk properties}

We can reproduce the observed SED of Par-Lup3-4 with a
single disk model (see Fig.\ref{sed_parlup34}).
The MCFOST model fits the overall shape of the SED  (see Fig.~\ref{sed_parlup34}), 
although it does not reproduce the 10 $\mu$m silicate feature (see subsection 5.3). 
  
  Figure~\ref{proba} shows that the most probable
  value of the inclination is $i$=81\degr. 
A Gaussian fitting to the probabilities
provides values of 76$\pm$9 degrees and 80$\pm$4 degrees for C03 and 2MASS, respectively.
These values  agree very well with the
inclination derived by \citet{FC2005} based on spectral lines related
to the stellar jet (78\degr$<i<$82\degr).

Our Bayesian analysis shows that the overall shape of the SED can only
be reproduced when including grains with a$_{\rm max}\ge$10~$\mu$m.
This result points towards dust processing within the disk of
Par-Lup3-4, as already observed in other low mass members of the same
association \citep[e.g. SST-Lup3-1, ][]{merin2007} or in other
associations of similar ages \citep[e.g.][]{Bouy2008,Luhman2007}.

For the inner disk radius, the best and most probable disk models
provide R$_{\rm in}\le$0.05~AU.  For the assumed dust composition and
disk geometry (from the best model), the constraint on the internal
radius corresponds to the distance where the temperature reaches the
sublimation temperature (T$\approx$1\,400~K).

Although fixed to 20\,AU, we have further investigated the outer disk
radius by comparing the synthetic images of the best MCFOST models
(both C03 and 2MASS) with the $K_s$-band NACO images (see
Section~2.1). The corresponding synthetic images were rebinned to the
NACO S27 camera pixel scale, and then convolved with a synthetic NACO
PSF obtained with the NAOS PS\footnote{NAOS Preparation Software}
software for the configuration and conditions of our observations, and
additional Poisson noise was added to match the signal-to-noise ratio
of the NACO image. The results are shown in the middle and right
panels of Fig.~\ref{nacoima}. The synthetic images are made of a
bright south-eastern nebulae and a faint north-western nebulae,
separated by a dark lane typical of high inclination disks. After
subtraction of a MOFFAT PSF on these synthetic images the faint
north-western nebulosity appears more clearly, as in the
PSF-subtracted NACO images, and align close to the disk direction.
Although the comparison of the observed and modeled data suggests the
disk is marginally resolved, more sensitive data is needed to confirm
this point.

The dust mass in the disk is only loosely constrained to M$_{\rm
  dust}$$<$10$^{-6}\,$M$_{\odot}$, that is, M$_{\rm
  disk}$$<0.1\,$M$_{Jup}$ assuming the typical gas to dust ratio of
100.  We have compared this value with the upper limit obtained from
the sub-mm data alone: assuming optically thin dust continuum
emission, we estimate the upper limit for the total mass (gas+dust)
from Par-Lup3-4 with

$$M = \frac{S_{870}~ D^{2}}{\kappa_{870} ~ B_{870}(T_{d})}\;$$

where $S_{870}$ is the flux density at 870$\mu$m, $D$ is the distance
to the source, $\kappa$$_{870}$ is the dust opacity per unit mass
column density at 870$\mu$m, and $B_{870}(Td)$ is the Planck function
at 870$\mu$m for a dust temperature $T_{d}$ \citep{Hil83}.  Adopting a
3$\sigma$ flux density upper limit of 13.5 mJy, a distance of 140\,pc
(see Section 5.2), a $\kappa_{870}$=0.015 cm$^{2}$ gr$^{-1}$
(following \citealt{Ose94}), and a $T_{d}$=15-20 K (see
\citealt{Tac07}), we derive an upper limit for the total mass of
1.9-3.0 $M_{J}$ depending on the adopted dust temperature.

Hence, the disk mass from our modeling is one order of magnitude
smaller than the upper limit obtained with APEX observations alone,
and is similar to that found in other disks around very-low mass
objects, e.g. SSM-Lup3-1 and 2M\,0438+2611, with masses of
$<$0.1\,M$_{Jup}$ and 0.3-0.6 M$_{Jup}$
\citep[see][respectively]{merin2007, Luhman2007}.

The surface density index shows a value closer to
$\alpha$$\sim$- 0.5 (only three values were covered by our analysis),
which is consistent with that found in disks around T Tauri stars
\citep{Dutrey1996,kitamura2002,Andrews2005}.

In the current state of the observations, we cannot derive any firm constraint 
on the flaring angle index, $\beta$, or the scale height $h_{0}$. However, we note
that the derived scale height for the 2MASS model, 1.28\,AU at 10\,AU, is consistent with the
hydrostatic scale height calculated from the modeled temperature structure of the disk (h$_{0}$ $\sim$1.05\,AU). 
This does not apply to the C03 model, with a derived scale height of 2.29@10AU and
a hydrostatic scale height of 1.04\,AU.

 The visual extinctions, $A_{\mathrm V}$, values are different
 depending on the near-IR dataset, 2.0$\pm$0.9 and 3.5$\pm$0.6\,mag
 for C03 and 2MASS data, respectively. In the case of the C03 model,
 the value is 3.6\,mag smaller than that derived using near-IR colors
 \citep[see][]{Comeron-2003}, a method that does not work properly in
 the case of edge-on disks.  For comparison, we have estimated the
 extinction caused by the cloud using an adaptive grid star-count
 method developed by \citet{Cambresy97}. We have used the 2MASS
 database instead of DENIS, since the former increases the number of
 objects in $\sim25\%$, doubles the spatial resolution and deepens
 $\sim 0.8$\,mag in the {\em J}-band.  From the sample of 2MASS
 objects in the region of interest, we have rejected those classified
 as Lupus~III members or candidate members
 \citep[][]{Wichmann1997a,Wichmann1997b,Krautter1997,Comeron-2003,
   Lopez2005,Gondoin2006} and those having {\it Hipparcos} distances
 lower than 100\,pc to avoid foreground objects. We have chosen $x =
 20$ and a $\sim 1.5 \times 0.8$ squared degrees field outside the
 cloud to derive the relationship between the local projected density
 and the limiting magnitude assumed in an unobscured area. The derived
 extinctions in the direction of Par-Lup3-4 are $A_{J}$=1.53,
 $A_{H}$=0.94, $A_{Ks}$=0.87, which translates into $A_\mathrm{V}$=5.4
 \,mag, following \citet{Rieke85}. The difference between this value
 and those coming from our modeling might suggest that the object is
 not placed at the most distant edge of the region but closer to the
 center of the cloud. This assumes that the C03 data represents the
 quiescent state of the source. However, a proper optical and infrared
 monitoring of Par-Lup3-4 is needed to identify this quiescent state
 and estimate the visual extinction associated with it.

The scaling factor, {\em q}, is smaller than 1.0 (we
obtain a value of 0.7 for the two models, placing the object at
98\,pc). This can be the result of a combination of uncertainties, in
particular, an error in the assumed distance, in the age (the star
can be younger than 1\,Myr old), and the error of the stellar radius
derived using evolutionary models.

Finally, our analysis of the two near-IR datasets illustrates the strong
effect of variability on the SED modeling. While the disk inclination
and the maximum grain size do not depend on the near-IR dataset used,
other parameters such as the scale height or the flaring angle are
significantly affected by variability. The SED modeling based on a single-epoch 
photometric dataset should therefore be interpreted with
caution. The degeneracy related to variability can be partially
addressed by using simultaneous observations at all wavelengths, but
also by a better coverage of the electromagnetic spectrum, and by spatially
resolved images of the disk at various wavelengths.

\subsection{The disk model and  the IRS spectrum of Par-Lup3-4}

We can reproduce the overall shape of the SED of Par-Lup3-4 with a single-disk model at
an inclination of $\sim$ 80\,degrees, in  good agreement with the value found by \citet{FC2005}.
Our model cannot reproduce the observed silicate feature, but this is not surprising because, as explained before,
we did not make a special treatment of the feature and used the  dielectric constants  from \citet{Mathis89}, 
which have a small silicate emission. 

Although the presence of a close to  edge-on disk can explain most of the observed properties of
Par-Lup3-4  (e.g. underluminosity, double-peak SED), it  is inconsistent with  the detection
of the silicate feature in emission.  At such a high inclination 
the feature is expected to be in absorption because the light emitted by the disk is scattered to the observer and the strong peak in the scattering cross-section of grains is obliterated by an even stronger absorption feature in the albedo, even for pure silicate grains.
    
Although the detection of the 10$\mu$m silicate feature in emission in Par-Lup3-4 is unexpected,
it was recently reported in several close to edge-on disks studied by \citet{merin2010}. The puzzling presence of this feature could be related to a more complex disk structure (e.g. with asymmetries and inhomogeneities) than the one assumed in 
theoretical models, or even to a presence of an optically thin (unlike those of embedded protostars) residual envelope.
Spatially resolved observations are needed to explore any of these possibilities.

\section{Conclusions}

We have analyzed new infrared and radio observations of the very
low-mass object Par-Lup3-4.  We have modeled its SED and characterized
its circumstellar environment.  Our main results can be summarized as
follows:

\begin{enumerate}

\item The SED of Par-Lup3-4 shows two emission peaks with a dip at 10
  microns, and a weak silicate feature in emission.  
 This shape can be interpreted as that of a star+disk
  system seen in a way that our direct line of sight is blocked by disk
  material, resulting in scattered light reaching the observer at all
  wavelengths shortward of 10 $\mu$m.  The presence of this close to
  edge-on disk can explain the apparent under-luminosity of the
  target. 

\item High angular infrared observations do not reveal extended
  emission from a thick envelope, but a possible marginal detection of
  the disk in scattered light.  This, and the non-detection of the 9.8$\mu$m
  silicate feature in absorption,
  suggests that Par-Lup3-4 is not a Class~I, but a Class~I/II or
  Class~II object.

\item A detailed Monte Carlo modeling of the SED has allowed us to
  constrain some of the disk properties, in particular it allowed us to derive a most probable
  inclination of $\sim$81\degr, which agrees very well with
  previous estimates based on its associated jet. 
 As in the case of several close to edge-on disks studied
by \citet{merin2010}, we detect silicates in emission
 at a high disk inclination. This might be related to a more complex disk structure
 that includes e.g. asymmetries and inhomogeneities. 

\item The maximum grain size, at least $>$10 $\mu$m, suggests an
  advanced level of dust processing as already observed in other low
  mass members of the same association \citep[e.g. SST-Lup3-1,
  ][]{merin2007} or in other associations of similar ages
  \citep[e.g.][]{Bouy2008,Luhman2007}.

\item Stellar variability affects the results of our SED modeling,
  which provides different results for some of the disk parameters,
  e.g. flaring angle or scale height, for different near-IR datasets.

\end{enumerate}

New measurements including deeper spatially resolved images of the
scattered light (e.g with $HST$), and far-IR and (sub)mm detections (e.g
with {\it Herschel}), would lead to better constraints on the disk properties
and on the formation of this very low-mass object. A more detailed analysis of
the {\it Spitzer} IRS spectrum would provide more information about the
origin and nature of the emission seen at 9.8$\mu$m.  Indeed, the presence of
this feature in emission at such a high inclination calls for more studies of mid-IR spectra of close
to edge-on disks, both on the theoretical and observational side. Finally, a
careful monitoring at optical, near-IR, and mid-IR wavelengths would
provide important information about the nature of the variability
(disk homogeneity, accretion, outflows) and its effect on SED
modeling.

\begin{acknowledgements}
 This research has been funded by Spanish grants MEC/ESP2007-65475-C02-02, MEC/Consolider-CSD2006-0070,
and CAM/PRICIT-S2009ESP-1496. MF was supported by the Spanish grants
 AYA2006-27002-E and AYA2007-64052. IdG is partially supported by
Ministerio de Ciencia e Innovaci\'on (Spain), grant AYA 2008-06189-C03
(including FEDER funds), and by Consejer\'{\i}a de Innovaci\'on, Ciencia y
Empresa of Junta de Andaluc\'{\i}a, (Spain).   C. Pinte acknowledges funding from the European Commission's seventh Framework Program as a Marie Curie Intra-European Fellow (PIEF-GA-2008-220891). The authors
 are very grateful to Angel Vicente for his support and help in using
 the computer resources at IAC. This research would not have been
 possible without access to the computing facilities provided by the
 Instituto Astrof\'\i sica de Canarias and the UC Berkeley Clustered
 Computing center at the Department of Electrical Engineering and
 Computer Sciences. The authors wish to acknowledge the contribution
 from Intel Corporation, Hewlett-Packard Corporation, IBM Corporation,
 and the National Science Foundation grant EIA-0303575 in making
 hardware and software available for the CITRIS Cluster, which was used
 in producing these research results. Finally, we acknowledge
 financial support from Programme National de Physique Stellaire
 (PNPS) of CNRS/INSU, France and Agence Nationale pour la Recherche
 (ANR) of France under contract ANR-07-BLAN-0221. This research has
 made use of the Condor\textregistered, developed and provided by the
 Condor Project at the University of Wisconsin Madison
 (http://www.condorproject.org/).

\end{acknowledgements}

\bibliographystyle{aa}
\bibliography{parlup}

\end{document}